\documentclass[12pt]{article}
\usepackage{amssymb}
\makeatletter
  
  \@addtoreset{equation}{section}
\makeatother
\def\bfr{\begin{flushright}}
\def\efr{\end{flushright}}
\def\si{\quad}
\def\sii{\qquad}
\def\noi{\noindent}
\def\vs{\vspace}
\def\vss{\vspace{.5cm}}
\def\begc{\begin{center}}
\def\endc{\end{center}}
\def\be{\begin{eqnarray}}
\def\ee{\end{eqnarray}}
\def\eq{\label}
\def\nn{\nonumber  \\}

\def\rarw{\rightarrow}

\def\lrarw{\leftrightarrow}

\def\a{\alpha}
\def\b{\beta}

\def\d{\delta}
\def\D{\Delta}
\def\e{\epsilon}

\def\z{\zeta}
\def\h{\eta}
\def\th{\theta}

\def\k{\kappa}

\def\L{\Lambda}
\def\m{\mu}
\def\n{\nu}
\def\x{\xi}

\def\o{\o}
\def\p{\pi}
\def\P{\Pi}

\def\t{\tau}

\def\f{\phi}
\def\F{\Phi}
\def\vf{\varphi}
\def\c{\chi}

\def\bra{\langle}
\def\ket{\rangle}
\def\fr{\frac}
\def\del{\partial}

\def\mxxb{\left( \begin{array}{cc}}           
\def\mxxe{\end{array} \right)}
\def\mxxxb{\left( \begin{array}{ccc}}         
\def\mxxxe{\end{array} \right)}
\def\mxxxxb{\left( \begin{array}{cccc}}       
\def\mxxxxe{\end{array} \right)}
\def\mxxxxxb{\left( \begin{array}{ccccc}}     
\def\mxxxxxe{\end{array} \right)}
\def\mxxxxxxb{\left( \begin{array}{cccccc}}     
\def\mxxxxxxe{\end{array} \right)}
\def\kakkob{\left\{ \begin{array}{c}}
\def\kakkoe{\end{array}  \right. }
\def\vecb{\left( \begin{array}{c}}
\def\vece{\end{array} \right) }

\def\real{\Bbb R}

\def\dx{\dot{x}}

\begin{document}

\bfr
TEP-09-01\\
March 2009\vs{2cm}
\efr

\begc
{\Large 
{\bf 
Relativistic Particle in Complex Space Time
}}\vs{2cm}

Takayuki Hori
\footnote{E-mail: hori\_{}takayuki@nifty.com}
 \vss

Department of Economics, Teikyo University, Hachioji, 192-0395 Japan\vs{2cm}

{\bf Abstract}\\
\endc

A modified version of the bilocal particle
is presented in terms of complex space time.
Unusual constraint structure of the model 
is studied, and a new concept of the physical equivalence is proposed in accordance with 
Dirac's conjecture.
It is found that in the quantum theory the physical state conditions are compatible with existence of eigenstates of momentum only when the dimension of space time is four.
An example of scattering amplitude is 
calculated.


\newpage
\setcounter{page}{1}
\pagestyle{plain}

\section{Introduction}
It has not been known why the dimension of 
space time is four.
A speculation was declared 
 for the possibility that in the  
eleven dimensional supergravity 
the dimension reduces to four 
for the sake of totally 
antisymmetric field strength of rank four\cite{FreundRubin}.
The idea of the spontaneous compactification 
opened  a vast area of investigations,
but  definite answers have not  
been known so far.

In the present paper we propose 
another approach for determining the 
dimension, where we only extends 
the ordinary space time to be spanned by complex valued coordinates.
Although a point particle in such a space 
has double  degrees of 
freedom corresponding to the real and the 
imaginary parts of its coordinates, 
some of which become unphysical because of 
an intrinsic gauge  symmetry.

This is motivated by the model of 
bilocal particle proposed by the author 
before a decade\cite{hori_1,hori_2}.
The model has gauge symmetry of $SL(2,\real)$, the anomaly free subalgebra of the Virasoro algebra  describing string.
In a word the bilocal particle is 
a tiny subsystem of string, 	
with classical behavior of a rigid 
stick, and has property of restricting the 
dimension of space time.
A detailed analysis of the model showed that 
there exist non-trivial  BRS cohomology classes  only when the dimension of space time is two or four\cite{hori_3}.
But the non-trivial physical sectors have non-zero ghost numbers, and the model 
cannot be treated in the usual quantization 
scheme.    

The proposed model in the present paper 
is a modified version of the bilocal model,
 and is free from the above difficulty.
The dimension of space time is fixed to be four by requiring the existence of  momentum eigenstates satisfying the physical state conditions.  

A particle in space time has complex 
valued coordinates, and is described 
by a Lagrangian defined in Section 3.
As will be shown there, the dynamics of 
the particle is so composed that it  enables one to handle 
the constraint algebra in an admissible  manner and a sufficient amount of physical content is constructed in the sector of no BRS ghost.

We also argue the concept of physical equivalence and the gauge degrees of freedom. 
Finally, we present  calculations of propagator and scattering amplitude of two complex particles, which 
may have counterparts in a possible 
field theory.


\section{Review of the bilocal model}

The bilocal model emerged through the 
following observations.
We start with the system of two independent and massless relativistic particles in $D$-dimensional space time, whose coordinates $(x^{\m}_1,x^{\m}_2)$ are functions of the internal {\it time} $\t$.
The action is 
\be
   I_2 =\int d\t\left(\fr{\dx^2_1}{2V_1}+\fr{\dx^2_2}{2V_2}\right),  \eq{action00}
\ee
where $V_1,V_2$ are einbeins (here and hereafter we suppress the space time suffices\footnote{The metric convention is $\h_{\m\n}={\rm diag}(-1,1,\cdots,1)$.}). The action is invariant under the reparametrization of each particle separately, which are expressed as
\be
 \d x_1 = \e_1\dx_1, \si \d x_2 = \e_2\dx_2,\si \d V_1 = \fr{d}{d\t}(\e_1V_1), \si \d V_2 = \fr{d}{d\t}(\e_2V_2),\eq{rep01}  
\ee    
where $\e_1$ and $\e_2$ are arbitrary functions of $\t$.
It has scarcely been pointed out that there exists another (global) symmetry of the action.
In fact, the transformation
\be
 \d x_1 = \fr{\e_0}{V_2}\dx_2, \si \d x_2 = \fr{\e_0}{V_1}\dx_1,\si \d V_1 =  \d V_2 = 0,\eq{sym00}  
\ee    
leaves action invariant, where $\e_0$ is arbitrary constant parameter. 
Of course the global symmetry  (\ref{sym00}) may have no relevance to any 
effects of physical significance, 
since the system is merely the two free  particles.

In refs.\cite{hori_1} it was found that the above symmetry is maintained if one adds extra term to the action, which mixes the coordinates of the two particles in a specific fashion. The modified action is
\be   I_{BL} =\int d\t\left(\fr{\dx^2_1}{2V_1}+\fr{\dx^2_2}{2V_2}+e(\dx_1x_2-\dx_2x_1)\right),  \eq{action_BL}
\ee
where $e$ is a constant with dimension of mass squared. We called the system a bilocal particle.
Apparently the added term in r.h.s of (\ref{action_BL}) spoils the symmetry of the direct product of reparametrizations of the two particles, but it turned  out that the very term not only recovers the  separate reparametrizations of two particles but brings forth  the global 
symmetry like (\ref{sym00}).
Explicitly the symmetry transformations are 
\be
 \d x_1 = \e_1\dx_1 +  \fr{\e_0}{V_2}\dx_2, && \sii \d x_2 = \e_2\dx_2 +  \fr{\e_0}{V_1}\dx_1,\eq{BLsym1}\\
\d V_1 = \fr{d}{d\t}(\e_1V_1) + 4e\e_0V_1,&& \sii \d V_2 = \fr{d}{d\t}(\e_2V_2) - 4e\e_0V_2,\eq{BLsym2}  
\ee    
where the infinitesimal parameters $\e_1,\e_2,\e_0$ are functions of $\t$, two of which are arbitrary, while another  is subjected to the constraint
\be
 \dot{\e}_0 + 2eV_1V_2(\e_2-\e_1) = 0. \eq{BLsym3}
\ee

Since the bilocal particle has two gauge invariance corresponding to $\e_1$ and $\e_2$, and the parameter $\e_0$ is determined by them through (\ref{BLsym3}) except its constant mode, the physical degrees of freedom would be the same as those of free two particles. 
The mixed term added to the action, however, gives rise to drastic changes on the system.
One striking feature is that the classical solutions to the Euler-Lagrange equations exhibit two moving particles with velocity of light, and their trajectories are completely parallel to each other\footnote{This fact has not been published in literature, but the similar results are demonstrated later in the present paper.}.
In other words the two particles are end points of a rigid stick moving with velocity of light. Moreover the forms of classical solution do not  depend entirely on $e$. 
These facts imply existence of some discontinuity of theories at $e=0$, where the system becomes two free particles moving in arbitrary directions.

The origin of the essentially global symmetry corresponding to $\e_0$ is 
traced back to existence of a tertiary first class constraint in the canonical theory, and it causes mismatch between  the number of gauge invariance and the number of first class constraints.
This  implies that the present model is an counter-example of Dirac's conjecture which claims that all first class constraints should generate gauge symmetries.

The reduction of the degrees of freedom in the classical solutions and the existence of tertiary constraint irrelevant to gauge symmetry are  pathological properties of the model, and the precise meaning of them has not been fully understood.
Another problem is in quantizing the system. In ref.\cite{hori_3}, the physical content of the model is investigated by calculating BRS cohomology classes.
It turned out that the physical states exist only in the sector of non-vanishing ghost number  in  specific dimensions of space time.
Because of these unusual properties of the model concrete calculations of physical quantities have not been demonstrated.


\section{Complex particle}

Let us start with the following action for the complex valued coordinates, $z^{\m},(\m=0,1,\cdots, D-1)$, 
\be
I_C = \int d\t \left(\fr{\dot{z}^2}{2V} + i\k\dot{z}\bar{z} + c.c.\right),
\ee
where $\dot{z}=\del z/\del\t$, and $V$ is complex valued einbein, and $\k$ is a real constant with dimension of mass squared.
The action describes dynamics of two real coordinates corresponding to the real and the imaginary parts of $z=x+ia$.
Depending on the value of $V$ the signature of the Hessian varies in such a way that some of the variables have wrong signs in the kinetic terms, {\it i.e.}, being ghosts. 
Thus the system contains unphysical components as well as physical ones, and we call the total system a complex particle.

The action is invariant under the following transformations,
\be
\d z = \e\dot{z} + \fr{\e_0}{\bar{V}}\dot{\bar{z}},
\sii
\d V = \fr{d}{d\t}(\e V) + 4i\k\e_0V,
\ee
where $\e$ and $\e_0$ depend on $\t$. While $\e$ has arbitrary complex value, $\e_0$ is real and subjected to the constraint,
\be
      \dot{\e}_0 - i\k V\bar{V}(\e - \bar{\e})=0.
\ee
Using the gauge freedom corresponding to $\e$, one can fix $z^0$.

The first integral of the Euler Lagrange(EL) equations is
\be
 \fr{\dot{z}}{V} + 2i\k(\bar{z}-\bar{c}) = 0,\sii \dot{z}^2 = 0, \eq{firstint}
\ee
with constant vector $c$.
We see that $z$ lies in a light cone.
If one chooses the gauge $z^0=e^{i\th}\t$,
with some angle $\th$, the einbein is determined by the $0$-component of (\ref{firstint}) as $V^{-1} = -2i\k e^{-2i\th}(\t - \t_0)$, where $z^0_0=e^{i\th}\t_0$.
Conversely if one fixes the gauge by this value of $V$, then the 
signs of the kinetic terms in the action change during  $\t$-evolution. 
A suitable choice for clearly distinguishing physical and ghost components is
\be
    V^{-1} = 2\k|\t-\t_0|.\eq{gauge1}
\ee
In this gauge choice the real and the imaginary parts of $z$ have correct and wrong signs, respectively, in the kinetic terms.

The EL equations are solved as follows.
Substituting (\ref{gauge1}) into the first equation of (\ref{firstint}) and differentiating it with respect to $\t$, we obtain a second order differential equation to $z$.
Then, putting the real part of $z$ as  $x(\t)=K(\t)(\t-\t_0)+x_0$, we get differential equation which can be solved for $K(\t)$ by integrations. Finally, substituting the solution back into (\ref{firstint}) we get
\be
   z(\t) = \left((\t-\t_0)k + \fr{w}{\t-\t_0}\right)(1+is(\t-\t_0))+ z_0,
\ee
where $s(\t)$ is the step function with value $1(-1)$ for $\t>0(\t<0)$, and $k$ and $w$ are $D$-dimensional real valued constant
vectors which are light-like and mutually orthogonal.
Note that, in the present gauge, $z^0(\t)$ approaches $\t + i|\t|$ asymptotically as $\t\rarw\pm\infty$.
Along the classical solution the spacial coordinates $x^i(\t)$ are expressed as functions of $x^0(\t)$,
which represent a gauge independent trajectory in the target space where $x^0$ is interpreted as {\it time}.
On the other hand the coordinates $a^i(\t)$ are double valued functions of $x^0(\t)$, though $a^i(\t)$ are single valued with respect to $a^0(\t)$.
Thus in the viewpoint of the target space the ghost particle moves, as time $x^0$ evolves,   along  trajectory completely parallel to that of $x^i$ until $\t<\t_0$, and moves away {\it after} then.
The ghost particle
could be observed in two points at a time $x^0$.

Now, the number of quantities whose initial values determine the later values\footnote{Here we use the words ``initial'' or ``later'' on account of $\t$.} of all dynamical variables through the equations of motion 
should coincide with the number of independent parameters appearing in the classical solution.
The latter  consists of  $2D$ coordinates of two particles at $\t=\t_0$ and $2D-3$ independent components of $k$ and $w$, which sum up to $4D-3$ degrees of freedom.
Thus $3$ variables, among $4D$ variables of coordinates and their time derivatives, should be redundant in determining the evolution of the system.
The reduction of the degrees of freedom is the key property of our model, and will be discussed later  in the canonical theory.

\section{Physical equivalence}

In order to clarify the gauge structure of the model let us describe the canonical theory.
The canonical momenta of $z$ and $\bar{z}$ are
\be
  p = \fr{\dot{z}}{V}+i\k\bar{z},\sii \bar{p} = \fr{\dot{\bar{z}}}{\bar{V}}-i\k z,\eq{momenta}
\ee
while those of $V$ and $\bar{V}$, denoting $\P$ and $\bar{\P}$, respectively, vanish.
Note that the momenta (\ref{momenta})
are {\it not} conserved quantities with respect to $\t$, since, by global translations of $z$,   Lagrangian varies by total derivatives, $i\k d/d\t(- \bar{z}\d z + z\d\bar{z})$.
Then the conserved momenta, denoted $\tilde{p}$ and $\tilde{\bar{p}}$, are obtained by subtracting the corresponding amounts from $p$ and $\bar{p}$, and we get
\be
    \tilde{p} = p +i\k\bar{z},\sii  \tilde{\bar{p}} = \bar{p} -i\k{z},
\ee
while the angular momentum defined by $M_{\m\n} = z_{[\m}p_{\n]} + \bar{z}_{[\m}\bar{p}_{\n]}$ is conserved.

Now the total Hamiltonian generating $\t$ development is
\be
    H_T &=& V\c_1 + \bar{V}\c_{-1}+v\P +\bar{v}\bar{\P},
\ee
where
\be
    \c_1 = \fr12(p - i\k \bar{z})^2,\sii  \c_{-1} = \fr12(\bar{p} + i\k z)^2,
\ee
and $v$ and $\bar{v}$ are the 
Dirac variables corresponding to the primary constraints, $\P\sim\bar{\P}\sim 0$.
The preservation of the primary constraints requires the secondary constraints, $\c_1\sim\c_{-1}\sim 0$, and the preservation of them requires the tertiary constraint
\be
  \c_0 = \fr12(p - i\k\bar{z})(\bar{p} + i\k z) \sim 0.
\ee
These constraint functions form a $SL(2,R)$ algebra with regard to Poisson brackets:
\be
    \{\c_n,\c_m\} = -2i\k(n-m)\c_{n+m}.\sii (n,m=0,\pm 1)
\ee

Since the total Hamiltonian does not contain the tertiary constraint function, $\c_0$, there appears a subtle problem 
 concerning to Dirac's conjecture.
The problem has been discussed by some authors in a slightly different contexts
\cite{Sugano,Frenkel}. 
There seems some disagreements in the 
authors, and one of the reasons  
may be  lack of common  definition of the concept of physical equivalence. 
We will examine this problem shortly.

Now some of the canonical variables, $z,p,,V,\P$, and their complex conjugates, are unphysical.
First of all the einbeins, $V$ and $\bar{V}$, are unphysical, since the time derivatives of them equal to the (unphysical) Dirac variables. To fix the einbeins and the Dirac variables amounts to fix the gauge, and it uniquely determines the time development of dynamical variables with an initial condition.
On the other hand the initial condition must be so determined that 
the primary constraints, $\P\sim\bar{\P}\sim 0$, are preserved through time development.
This requirement is satisfied if and only if one chooses such an initial condition for the canonical variables that the secondary and tertiary constraints are all satisfied at the initial time.

According to Dirac\cite{Dirac0}, one may say that two points in the phase space is physically equivalent if the two points are reached through the canonical equations of motion, which are not unique due to the gauge freedom, from a common initial point in the phase space.
Transformation from a point in the phase 
space to the physically equivalent point 
is called gauge transformation.

Let us consider the transformations of a canonical variable $q$, generated by the constraint functions $\c_{\pm 1},\c_0$ and $\P,\bar{\P}$;
	\be
     \d q = \{q,Q\},\sii Q = \sum_{a=0,\pm 1}\!\!\e_a\c_a + \x\P+\bar{\x}\bar{\P},
\ee
where transformation parameters are time dependent with $\e_a(0)=\x(0)=0$.
If time development of $q$ is generated by the total Hamiltonian, {\it i.e.}, $\dot{q}= \{q,H_T\}$, then  it turns out, using the Jacobi identity, that $\tilde{q}=q+\{q,Q\}$ develops as
\be
   \dot{\tilde{q}} = \{q,\tilde{H}\}\big|_{q=\tilde{q}}+O(\e^2),\sii\sii\sii\\
   \tilde{H} = H_T+\tilde{Q}+\dot{\x}\P+\dot{\bar{\x}}\bar{\P},\sii
    \tilde{Q} = \sum_{a=0,\pm 1}\dot{\e}_a\c_a +  \{Q,H_T\}. 
\ee
If one chooses the parameters $\e_a$ such that $\tilde{Q}$ vanishes, then $\tilde{H}$
 and $H_T$ are different only by the Dirac variables, and the  transformations, with suitable  redefinition of parameters, precisely coincide with (\ref{BLsym1}) and (\ref{BLsym2}) in the Lagrangian theory.
Physical equivalence in the canonical theory, however, is the wider concept than that of Lagrange theory.
 The two quantities $q$ and $\tilde{q}$ are physically equivalent, in the canonical theory, if the difference between $\tilde{H}$ and $H_T$ is absorbed into redefinition of the unphysical variables $V$ and $\bar{V}$ and of the Dirac variables.
This is not the case in our model, since there does not exist $\c_0$ term in the Hamiltonian, indicating breakdown of Dirac's conjecture.

What is the origin of the reduction of the degrees of freedom mentioned in the previous section?
The key element is in the definition of the concept of the physical equivalence.
As mentioned before Dirac defined that two points in the phase space are physically equivalent if they are obtained through time evolution from a common initial condition.
In the present paper we postulate another definition of the physical equivalence represented by the following proposition:\vss

\noi {\bf Proposition:} Two points in the physical phase space are  physically equivalent if {\it all} conserved quantities take the same values at the two points.\vss

Obviously any  two points in the physical phase space, lying in an orbit satisfying the equations of motion, have the same conserved quantities.
Conversely, if one can derive the existence of 
such an orbit connecting two points having the same values of all conserved quantities, 
then the concept of the physical equivalence defined above is the same as that of Dirac.

In our model the maximum set of conserved quantities are $\tilde{p}, \tilde{\bar{p}}, M_{\m\n}$ and $\c_0$.
The last one is the Noether charge 
corresponding to the global symmetry 
of $\e_0$, which must be zero along  physically admissible orbits.
The transformations generated by the  constraint functions, $\c_a,(a=0,\pm1)$, transform a point in the phase space into the physically equivalent point in the sense of the above proposition, since they keep the conserved quantities defined above invariant.
Therefore Dirac's conjecture holds in the 
sense of our definition of the physical 
equivalence.

As an illustration we show 
that our definition of the physical equivalence is equivalent to that of Dirac in the case of relativistic particle. The proof is as follows.
The action  and the Hamiltonian of relativistic particle are defined by
\be
  I=\int d\t\fr{\dot{x}^2}{2V},\sii H= \fr12 Vp^2 + v\P_V,
\ee
where $V$ is the einbein and $p$ and $\P_V$ are momenta conjugate to $x$ and $V$, respectively.  The general solution to the canonical equations of motion is
\be
&&
x^i(\t) = \fr{k^i}{k^0}x^0(\t) + y^i,
\sii
x^0(\t) = f(\t),
\sii
V(\t) = \fr{\dot{f}}{k^0},
\eq{rpsol1}
\\
&&
p_{\m}(\t) = k_{\m},
\sii
\P_V(\t) = -\fr12 k^2\t + \P_V(0),
\sii 
v(\t) = \fr{\ddot{f}}{k^0},
\ee
where $k^i,y^i$ are arbitrary constants and $f$ is an arbitrary differentiable function of $\t$.
Suppose that at two points A and 
B in the phase space, whose coordinates are $(x_1,p_1)$ and $(x_2,p_2)$, respectively, the conserved quantities, $M_{\m\n}=p_{[\m}x_{\n]}$ and $p_{\m}$ take the same values.
Now, let's constitute the trajectory which passes through A and B at $\t=\t_0$ and $\t=\t_1$, respectively.
Note the trajectory defined by
\be
  x^i(\t_0)=x^i_1,\sii f(\t_0)=x^0_1,
\ee
passes through A at $\t=\t_0$.
Since $M_{0i}$ takes the same value at A and B, {\it i.e.}, $x^i_1-k^ix^0_1/k^0=x^i_2-k^ix^0_2/k^0$(here the common value of $p$ is denoted by $k$), we have
\be
x^i(\t_1)
=
\fr{k^i}{k^0}(f(\t_1)-x^0_2)+x^i_2.
\ee
Because $f(\t)$ is arbitrary we can choose it such as $f(\t_1)=x^0_2$.
Then, $x^i(\t_1)=x^i_2$, and we see that $x^i(\t)$ is the solution passing through B.
If B and B${}'$ in the phase space take the same values of $M_{\m\n}$ and $p_{\m}$, then as is shown above there exist solutions of the equations of motion, one of  which combines A and B, and another combines A and B${}'$.
That is, B and B${}'$ are physically equivalent in Dirac's sense.
As expected the equivalence classes are nothing but the {\it trajectories} moving in $(D-1)$-dimensional space with velocity of light with regard to {\it time}, $x^0$.


In order to prove, if possible, that  the physical equivalence defined above implies Dirac's equivalence, it is required to get general solution to the equations of motion in a general gauge as demonstrated above in the relativistic particle. This may be  rather hard task in our model, since the equations reduce to 2nd order differential equations with general coefficient functions.
In the present paper we assume that the concept of physical equivalence proposed above is meaningful as a first principle or by definition.
It is probable that the origin of the reduction of the degrees of freedom found in the classical solution to the equations of motion is in the presence of the {\it gauge} symmetry generated by $\c_a,(a=\pm 1,0)$. 
    

\section{Quantization}

In this section we present the quantum theory, 
assuming that our model is a constrained 
Hamiltonian system with gauge symmetries 
generated by $\c_a,(a=0,\pm 1)$.
This point of view is consistent with the reduction of the classical degrees of 
freedom mentioned in Section 3.

Let us represent the dynamical variables 
as linear operators on the space 
of differentiable functions of $z$. 
The classical constraint functions are replaced by the following operators:
\be
    L_{1} &=& \fr{1}{4\k}(-i\del - i\k \bar{z})^2,\\
    L_{-1} &=& \fr{1}{4\k}(-i\bar{\del} + i\k {z})^2,\\
    L_{0} &=& \fr{1}{4\k}(-i\bar{\del} + i\k {z})(-i\del - i\k \bar{z}) + \a,\eq{order_a}
\ee
where $\del=\del/\del z$, and the constant $\a$ represents the ambiguity due to the operator ordering.
The above operators constitute a basis of $SL(2,\real)$ with central term as
\be	
 [L_{n},L_{m}] = (n-m)\left(L_{n+m} + \left(\frac{ D}{4}-\alpha \right)\delta _{n+m}\right),\sii  (n, m = 0, \pm 1), \eq{sl2r}
\ee 
where $D$ is the dimension of space time.

According to the gauge algebra (\ref{sl2r}) the BRS charge is defined by
\be
    Q = \sum_{n=0,\pm 1}c_{n}L_n  -  \fr{1}{2}\sum_{n,m=0,\pm 1}(n-m)c_{n}c_{m}\fr{\del}{\del c_{n+m}},
\ee
where $c_a,(a=0,\pm 1)$ are the BRS ghost variables.
The square of the BRS charge is 
\be
Q^2
=2
\left(
      \fr{D}{4}-\a
\right)
c_1c_{-1}.
\ee
As in the ordinary gauge theory we 
require the nilpotency of $Q$ so that
the ordering ambiguity is fixes as $\a=D/4$, which also eliminates the central term in (\ref{sl2r}).

In the classical theory  the
 constraints, $\c_n=0,(n=0,\pm 1)$, are imposed along the physically admissible orbits.
In the quantum theory we cannot regard them as operator equations because they have no solution, and 
the  conditions are relaxed 
 so that 
a product of the constraint operators  has vanishing matrix elements between any physical states, $|\vf\ket$ and $|\f\ket$:
\be
\bra\vf |L_{n_1}\cdots L_{n_N}|\f\ket=0.
\eq{psc1}
\ee
This  is realized  by requiring 
\be
   L_1|\f\ket =L_0|\f\ket = 0,
\eq{psc2}
\ee
for physical state $|\f\ket$, since we have $\bra\f |L_{-1}=0$ 
by virtue of the hermiticity,  $L^{\dagger}_{1}=L_{-1}$, the property  lacking  in the original bilocal model.
The above conditions for physical states 
are analogous to those of string model, 
and seems most natural ones.

In order that our model is physically 
meaningful 
 there should exist the eigenstates of 
momentum.
As is shown shortly this requirement gives rise to restriction on the space time dimension.
The conserved quantities derived by 
the invariance under the space time translations are
\be
\tilde{p} = -i\del + i\k\bar{z},
\sii 
\tilde{\bar{p}}= -i\bar{\del} - i\k z.
\ee
Thus the momentum should be combinations of these quantities. From the reality of 
eigenvalues, it should have the form
$
P=\b\tilde{p}+\bar{\b}\tilde{\bar{p}},
$
with arbitrary complex constant $\b$.
Any two of the eigenstates of $P$ 
would be taken as independent momentum eigenstates.
However, we regard one of them as the physical momentum state, since these two operators are not 
mutually commuting and have not simultaneous eigenvalues.
Without loss of generality we can put $\b=1$, since  global rotations can change $\b$.

The eigenstates of $P$ with $\b=1$, which satisfy the physical state conditions, are obtained as follows.
Let us write the eigenstate of $P$ with
eigenvalue $k$ as
\be
   |k\ket = e^{-\k z\bar{z}}f(z,\bar{z}).
\ee
By the physical state condition $L_1|k\ket=0$ we see
that $f(z,\bar{z})$ is an harmonic function with respect to $z$.
The eigenvalue equation $P|k\ket = k|k\ket$ reduces to
\be
(\del + \bar{\del} 
- 2\k\bar{z} - ik)
f(z ,\bar{z})=0.\eq{PeigenEq}
\ee
This equation is of the form with separate
variables, and has the solution
\be
f(z,\bar{z}) = e^{ik_1z + i(k-k_1)\bar{z} + \k \bar{z}^2},\eq{fzz}
\ee 
with arbitrary separation constant $k_1$.
Multiplying arbitrary function $a(k_1)$ to
(\ref{fzz}), and integrating over $k_1$,
we obtain the general solution to 
(\ref{PeigenEq}) as
\be
f(z,\bar{z})
= e^{ik\bar{z}+ \k\bar{z}^2}g(z - \bar{z}),  \eq{fzz2}
\ee
where
$g$ is an arbitrary differentiable function which can be Fourier expanded.
Since $f(z,\bar{z})$ is harmonic with respect to $z$, $g(y)$
must be an harmonic function.

Finally, the condition $L_0|k\ket=0$ 
reduces to
\be
(\bar{\del}\del -2 \k{z}\del - 4\k\a)
f(z,\bar{z}) =0.
\ee
Substituting (\ref{fzz2}) into this,  we get
\be
  \left[
     \left(
        y^{\m} - \fr{ik^{\m}}{2\k}
     \right)\fr{\del}{\del y^{\m}}
      + 2\a
  \right]
  g(y)
= 0,
\ee
with the solution 
$g(y) \propto [(y - \fr{ik}{2\k})^2]^{-\a}
$.
Since $g$ is harmonic, we see
\be
\del^{\m}\del_{\m}g
\propto
\fr{
      D-2(\a+1)
   }{
      \left[\left(
           y - \fr{ik}{2\k}
       \right)^2\right]^{\a + 1}
   }
=0.
\ee
Along with the nilpotency
condition of the BRS charge, we see
$D/2-1 = \a=D/4$.
Hence we find that the momentum eigenvalue equation is compatible to the physical state conditions only when $\a=1$ and $D=4$. This result does not depend on the 
choice of $\b$ in the definition of the 
momentum.
The physical eigenstate of the momentum 
in four dimensions is written as
\be
|k\ket \propto  
\fr{
       e^{ 
           ik\bar{z}
          + \k\bar{z}(\bar{z}-z)
         }
    }{
       \left(
            z - \bar{z} 
            - \fr{ik}{2\k}
       \right)^2
    } \equiv \F[k](z,\bar{z}).
\eq{state_k}
\ee

There are {\it supurious}  states defined by $L^n_{-1}|k\ket,(n=1,2,..)$, which are  orthogonal to all physical states and have zero norm.
In the string theory there are 
many supurious states which are physical 
and have zero norm, especially in the 
critical dimension. Existence of these 
states in the string theory suggests some  underlying gauge 
invariance, since they must be decoupled from physical $S$-matrix. In the present model, however, supurious states are all  unphysical by virtue of the constraint 
algebra without central term, and do not 
enter in physical $S$-matrix from the outset.


\section{Propagator and amplitudes}

In order to build a  model of
particle physics based on the complex 
space time proposed in the present 
paper, it may be most promising to 
construct the field theoretic formulation.
To speculate the form of the possible field theory, let us calculate some quantities containing  functional integrations, which would 
have counterparts in the field theory.

It is well known that the
quantity
\be
\int_{0}^{\infty} \!\!\!dt
\int \!\!d\m_x~
e^{ik_1x_1-ik_2x_2}
\exp{    
     \int_0^{t}d\t
     \left(
             \fr{\dot{x}^2}{2V} - \fr12 m^2V
     \right)
    } 
\ee
is proportional to the propagator
in the scalar field theory with
mass $m$,
where $x$ is the coordinates of 
ordinary relativistic particle with 
boundary values, $x(0)=x_1,~x(t)=x_2$, 
and $V$ is the (constant) einbein and $d\m_x$ is 
a suitable functional measure.
Analogously it may be worth while 
to calculate the following quantity
in our model:
\be
P(k_1,k_2)
=
\int_{0}^{\infty} \!\!\!dt
\int \!\!d\m_z~
\F_1\F^*_2
\exp{
  i\int_{0}^{t}\!\!\!d\t
   \left(
       \fr{\dot{z}^2}{2V}
      +i\k\bar{z}\dot{z}
      +{\rm c.c.}
   \right)},
\eq{prpgt0}
\ee
where $\F_a=\F[k_a](z(\t_a)),(a=1,2),\t_1=0,\t_2=t$.

We calculate it in the gauge that $V$
is positive real constant.
Substituting (\ref{state_k}) into (\ref{prpgt0}) and
separating the functional integration 
to the real and the imaginary parts of
$z=x+ia$, (\ref{prpgt0}) is written 
as
\be
P(k_1,k_2)
&=& 
\fr{e^{\fr{k_1^2+k_2^2}{8\k}}}{16}
\int_{0}^{\infty} \!\!\!dt
\int_{-\infty}^{\infty}\!\!dx_1
e^{ik_1x_1}
\int_{-\infty}^{\infty}\!\!dx_2
e^{-ik_2x_2}
\int d\m_a
\nn
&&\sii
\times
  \fr{
     \exp{
        \left[
          -2\k
          \left(
          \left(
            a_1-\fr{k_1}{4\k}
          \right)^2
          +
          \left(
            a_2-\fr{k_2}{4\k}
          \right)^2
          \right)
        -i
          \int_{0}^{t}\!d\t
          \fr{\dot{a}^2}{V}
        \right]
      }
   }{
     \left(
        a_1-\fr{k_1}{4\k}
     \right)^2
     \left(
        a_2-\fr{k_2}{4\k}
     \right)^2
   }
\nn
&&\sii
\times
G[x_1,x_2,t;a],  \eq{P_1}
\ee
where
\be
G[x_1,x_2,t;a]
=
\int_{x(0)=x_1}^{x(t)=x_2} \!\!     d\m_x~
\exp{
  i\int_{0}^{t}\!d\t
      \left(
         \fr{
            \dot{x}^2
          }{
            V
          }
          +4\k a\dot{x}
      \right)
    }.
\ee
This is calculated in the usual method by
substituting
\be
1
=
\int d\m_p~
\exp{
  \left[
     -i\int_0^t
          d\t~ \fr{V}{4}
          \left(
             p - \fr{2}{V}
             (\dot{x}+2\k aV)
          \right)^2
  \right]
} \eq{unit}
\ee
into the functional integral,
where
$d\m_p = \prod_{i=1}^N\fr{iV^2\D\t^2d^4p(\t_i)}{16\p^2},~ \t_1=\D\t,~\t_2=2\D\t,...,\t_N=t
$.
Integrating in the order, 
$x(\D t)\sim x(t-\D t),~p(2\D t)\sim p(t)~, p(\D t),x_1,x_2$, and shifting
the integration variables as $a\rarw a+k/4\k$,  we get
\be
P(k_1,k_2) &=& \d(k_1-k_2)P(k_1),
\\
P(k)
&=&
Ne^{\fr{k^2}{4\k}},
\\
N
&\propto&
\int_{0}^{\infty} \!\!\!dt~
\int
d\m_a~
\fr{
  e^{-2\k(a_1^2+a_2^2)}
}{
  a_1^2
  a_2^2
}
\exp{
   i\int_{0}^{t}\!d\t
     \left(
        \fr{-\dot{a}^2}{V}
        -
        4\k^2V
        a^2 
     \right).
}
\ee
$N$ is a mere constant independent of $k$.
This result is not connected continuously to the limiting case $\k=0$, where 
we expect the behavior $\sim 1/k^2$.
The singularity at $k^2=0$ disappears by summing up all orders of 
$\k$ in the numerator, before the shift of the integration
variables, which in fact has the $1/k^2$
behavior in the lowest order of $\k$.
The characteristic feature of the
propagator may give some information
in the field theory.

As another example let us consider scattering process of two complex particles with initial and 
final momenta, $(k_1,k_2)$ and $(k_{-1},k_{-2})$, respectively.
We consider the following quantity:
\be
A(k_1,k_2,k_{-1},k_{-2}) 
=
\int_{0}^{\infty} \!\!\!dt
\int \!\!d\m_{z_1}d\m_{z_2}~
\F_1\F_2\F^*_{-1}\F^*_{-2}
\exp{i
  \int_{0}^{t}\!\!\!d\t
  (L_1 + L_2 + L_{\rm int}) 
  },  
\eq{amp_1}\nn
\ee
where
\be
\F_{\pm m}
=\F[k_{\pm m}](\z_{\pm m}),\sii
\z_{+m}=z_{m}(0),\sii 
\z_{-m}=z_{m}(t).\sii (m=1,2)
\sii
\ee
$L_m,(m=1,2)$ are the Lagrangian of
the respective complex particles, 
\be
L_m
=
\fr{
   \dot{z}_m^2}{2V_m}
   +i\k\dot{z}_n\bar{z}_m + {\rm c.c.},\sii (m=1,2)
\ee
and $L_{\rm int}$ describes an 
interaction of them.
It is expected that the {\it amplitude}
defined by (\ref{amp_1}) may correspond
to some of the Feynman diagrams in the
possible field theory.

The interaction term must be so chosen 
that the total momentum is conserved.
In our choice $\b=1$ in the definition
of the momentum, the requirement is
that $L_{\rm int}$ is invariant
under global translations of the
real part of $z_m=x_m+ia_m,(m=1,2)$.
A candidate of it is of the form 
$L_{\rm int}=f_1(x_1-x_2)f_2(\dot{x},a,\dot{a})$.
We take the simplest one
\be
   L_{\rm int} = 4g(x_1-x_2)(\dot{a}_1 - \dot{a}_2),
\eq{Lint}
\ee
where the constant $g$ has the dimension
of mass squared.
This is written in terms of $z$ as
\be
 L_{\rm int}
=
 -ig
(\dot{z}_1-\dot{z}_2)
(\bar{z}_1-\bar{z}_2)
-\fr{ig}{2}
\fr{d}{d\t}(z_1-z_2)^2 + {\rm c.c.}
\ee
The integrations in (\ref{amp_1}) over the constant mode of
$x_a,(a=1,2)$ will give rise to the delta 
function guaranteeing the conservation of the total momentum.

Calculation of (\ref{amp_1}) can be done
in the similar manner as that of 
propagator.
In performing the functional integration
with fixed boundary of $x_1(\t),x_2(\t)$,
one insert (\ref{unit}) to the
respective formula.
As a result we get
\be
A(k_1,k_2,k_{-1},k_{-2})
=
\d
\left(
   k_1+k_2-k_{-1}-k_{-2}
\right)
e^{
   \fr{1}{8\k}
   \sum_{n}
     k^2_n
 }
\int_{0}^{\infty} \!\!\!
dt~
T(k_1,k_2,k_{-1},k_{-2},t),
\nn
\ee
where
\be
&T&\!\!(k_1,k_2,k_{-1},k_{-2},t)
\nn
&\propto& 
\int
d\m_{a_1}
d\m_{a_2}
  ~\d^4\left(
      k_1-k_{-1}
    + 4g(
           \a_{-1} -  \a_{1}
         - \a_{-2} + \a_{2}
        )
    \right)
\nn
&&\sii
\times
\left(
  \prod_{n=\pm 1,\pm 2}
  \fr{
     \exp{-2\k
          \left(
            \a_n-\fr{k_n}{4\k}
          \right)^2
      }
   }{
     \left(
        \a_n-\fr{k_n}{4\k}
     \right)^2
   }
\right)
\exp{-i
    \sum_{m=1}^2
     \int_{0}^{t}\!d\t
     \left[
        \fr{\dot{a}^2_m}{V_m}
     +
        4\k^2V_m
          \left(
            a_m - \fr{k_m}{4\k}
          \right)^2
     \right]
},
\nn
\\
&&
\!\!\!\!\!\!
\a_n = {\rm Im}\z_n.\sii (n=\pm 1,\pm 2)
\nonumber
\ee
The total momentum is conserved as expected.
As in the calculation of the propagator
one shifts the integration variables
by $a_m(\t)\rarw a_m(\t)+k_m/4\k,(m=1,2)$,
 where $k_m,(m=1,2)$ are the initial momenta of the two complex particles.
In this case, however, the momentum 
dependence does not disappear, since
the formula contains the final variables
as $\a_{-m}-k_{-m}/4\k,(m=1,2)$, with $k_m\ne k_{-m}$.
Suitably rescaling time variable $t$ 
we get the following expression with 
parameter integrals:
\be
&T&\!\!(k_1,k_2,k_{-1},k_{-2},t)
\nn
&\propto& 
\left(
  \prod_{m=1}^2
  \int_{i}^{i + \infty}\!\!\!\!\!\!
  d\k_m
  \int_{i}^{i + \infty}\!\!\!\!\!\!
  d\k_{-m}
  \int
d^4\a_m
d^4\a_{-m}
\right)
g^{-4}
\d^4	\left(
        \a_{-1} -  \a_{1}
      - \a_{-2} +  \a_{2} - \fr{\D k_1}{4g}
        \right)
\nn
&&
\times
\prod_{m=1}^2
\exp{2\k i
    \left\{
        \k_m
        \a_m^2
        +
        \k_{-m} 
        \left(
            \a_{-m} - \fr{\D k_m}{4\k}
        \right)^2
     \right\}
    }
K(\a_m,\a_{-m},t),
\eq{amp_3}       
\ee
where $\D k_m=k_{-m}-k_m$ is the 
momentum transfer of $m$th complex 
particle, and
\be
K(\a,\b,t)
=
\fr{1}{\sinh^4{t}}
\exp{
     \left\{-8\k i
     \fr{
          2\a\b 
          -
          (\a + \b^2)
          \cosh{t}
        }{
          \sinh{t}
        }  
     \right\}     
    }       
\ee
is the well known kernel of the
harmonic oscillator\cite{FeynmannHibbs} (with wrong sign) in four dimensions.
(Here we choose the gauge that $V_1=V_2$.)

Note the expression (\ref{amp_3}) is  symmetric under $1\lrarw 2$.
One of the space time integrations over
$\a_n,(n=\pm 1,\pm 2)$ is carried out
by the delta function, leaving 
exponential of bilinear terms of $\a_n$.
If one chooses $\a_1$ as that variable and 
carries out the remaining space time integrations, one gets the following 
expression of the amplitude without manifest symmetry 
under $1\lrarw 2$:
\be
T(k_1,k_2,k_{-1},k_{-2},t)
\propto 
g^{-4}
\left(
  \prod_{m=1}^2
  \int_{i}^{i + \infty}\!\!\!\!\!\!
  d\k_m
  \int_{i}^{i + \infty}\!\!\!\!\!\!
  d\k_{-m}
\right)
\fr{
   \exp{\left\{
       \fr{
            i\k\D k^2
           }{
            2g^2
           }
       (
         -vM^{-1}v
         +\L
        )
       \right\}}
    }{
       \sinh^{16}{t}
       (\det{M})^8
    },
\eq{amp_4}
\nn
\ee
where
\be
M
&=&
\mxxxb
   \k_1+\k_{-1}-8T_{-}
 & 
  -\k_1+4T_{-}
 &
   \k_1-4T_{-}
\\
  -\k_1+4T_{-}
 &
   \k_1 + \k_{-2} + 8T
 &
   \k_1-4T_{+}
\\
   \k_1-4T_{-}
&
   \k_1-4T_{+}
&
   \k_1+\k_2 + 8T 
  \mxxxe,\si
\\
v
&=&
\vecb
-\k_1-\fr{g}{\k}\k_{-1}+4T_{-}\\
\k_{1} + \fr{g}{\k}\k_{-2}+ 4T\\
-\k_{1}+ 4T
\vece,
\sii
\L 
=
\fr{\k_1}{4}  + 
\fr{g^2}{4\k^2}(\k_{-1}+\k_{-2})
+ T,
\ee
\be
T_{\pm}
= 
\fr{
     1\pm\cosh{t}
    }{
     \sinh{t}
    },\sii
T
=
     \coth{t}.\sii
\ee
Since the  integrations over parameters in the 
above formula do not give simple
expressions in terms of elementary 
functions,
we give up further demonstration of calculations.

It would be expected that the amplitude 
possesses the property like duality as 
a consequence of the underlying $SL(2,\real)$ gauge symmetry.   
This is not the case because the 
interaction term violates the symmetry 
 except in the case $\k=g$, where 
we can verify the existence of an enlarged gauge symmetry.
However, even if $\k=g$, we are not able to see  such  property, since our result (\ref{amp_4}) is expressed in an asymmetric form under $1\lrarw 2$.

\section{Conclusion}

We have proposed 
a model of complex particle which is
a modified version of the bilocal 
model proposed by the author.
Our model may be interpreted in a sense as complexification of the ordinary space time coordinates without introducing 
 extra degrees of freedom as far as possible.
In other words the reality of the coordinates is relaxed in such a way  
that only the physical content of the space time is kept real.
The real and the imaginary parts of 
the complex coordinates and their time 
derivatives are related to one another 
according to the constraint structure 
of the starting Lagrangian.

The canonical theory  reveals some unusual properties of the complex 
particle.
The most problematic one is the mismatch 
of the numbers of first class constraints 
and the gauge degrees of freedom, {\it i.e.}, the breakdown of Dirac's conjecture.
We have argued that if one redefines the concept of physical equivalence as 
stated in Section 4, then the first class constraints can be used to classify the physically equivalent points in the physical phase space. 
This definition is equivalent to Dirac's one in the case of the ordinary relativistic particle.

In the quantum theory we set  natural 
conditions for a physical state in the 
analogous form as string, and have solved 
them in the sector of ghost number zero. 
The real momentum is defined to be 
one of the two momenta corresponding to the real and the imaginary coordinates 
of a complex particle 
or a linear combination of them, 
which should have eigenstate in 
the space of physical state vectors.
A consequence is that the 
dimension of the space time is restricted 
to be four.

In order to get some insights into the 
possible field theory we have calculated 
propagator and amplitude in the first 
quantized scheme.
It turned out that the expected $1/k^2$ behavior of the propagator 
 in all orders of $\k$ sums up to disappear, and the amplitude does not 
indicate any striking symmetry.

Unlike string the complex particle has 
not apparent internal degrees of freedom,
 and may not be a basis of unified theory 
 of elementary particle at least in the 
 present form.
Although the model is in a premature stage,  
we hope it will open a new area of particle 
physics, whereas various field theories 
could be reconstructed in a new perspective.

\end{document}